# ROSAT PSPC observation of the X-ray faint early-type galaxy NGC5866

S. Pellegrini

ESO, Karl Schwarzschild Str.2, D-85748 Garching bei München

**Abstract.** We present the results of the analysis of the ROSAT PSPC pointed observation of the S0 galaxy NGC5866. Previous *Einstein* observations had revealed that this galaxy has a low X-ray to optical ratio $L_X/L_B$. Theoretical models of the X-ray emission of early-type galaxies had suggested that in objects of this kind the emission is not dominated by the presence of a hot diffuse gas, but should originate in stellar sources. We discuss the evidence in favor of this hypothesis following from the results of the analysis of the ROSAT PSPC data. The X-ray spectrum shows properties similar to those of the other two low $L_X/L_B$ early-type galaxies observed by ROSAT so far, including the presence of very soft emission. So, these galaxies can be recognized as a group with homogeneous properties, and a more exhaustive picture of the nature of the X-ray emission across the $L_X$–$L_B$ plane of early-type galaxies can be outlined.

We also discuss the importance of rotation in the X-ray emission of S0 galaxies, and suggest that it could explain why on average S0s are less X-ray luminous than ellipticals.

**Key words:** Galaxies: elliptical and lenticular, cD - Galaxies: individual: NGC5866 - Galaxies: ISM - X-rays: galaxies - Radiation mechanisms: miscellaneous

## 1. Introduction

After the launch of the *Einstein* satellite (Giacconi et al. 1979), it was first realized that normal early-type galaxies are X-ray emitters, with 0.2–4 keV luminosities ranging from $\sim 10^{40}$ to $\sim 10^{43}$ erg s$^{-1}$ (Fabbiano 1989; Fabbiano, Kim, and Trinchieri 1992). The X-ray luminosity $L_X$ is found to correlate with the blue luminosity $L_B$, but there is substantial scatter at any fixed $L_B > 3 \times 10^{10} L_\odot$ (Fig. 1). The observed X-ray spectra of the brightest objects are consistent with thermal emission from a hot, optically thin gas, at a temperature of $\sim 1$ keV, accumulated from the stellar mass loss within the galaxies (Canizares, Fabbiano and Trinchieri 1987).

The scatter in the $L_X$–$L_B$ diagram of early-type galaxies has been explained in terms of environmental differences (i.e., varying degrees of ram pressure stripping due to the interaction with the intracluster or intragroup medium, White and Sarazin 1991), or in terms of different dynamical phases for the hot gas flows, ranging from winds to subsonic outflows to inflows (Ciotti et al. 1991). In the latter interpretation, at any fixed $L_B$ X-ray faint galaxies are still in the wind phase, and the emission is mostly accounted for by stellar sources; in the bulk of galaxies the gas flows are in the outflow phase; in the X-ray brightest galaxies the hot gas dominates the emission, being in the inflow phase. The correlation of X-ray and optical luminosities is seen as a consequence of the fact that optically brighter galaxies are characterized by a larger binding energy per unit mass, i.e., as an implication of the Faber-Jackson (1976) $L_B - \sigma$ relation (where $\sigma$ is the central stellar velocity dispersion).

The knowledge of the contribution of stellar sources to the total X-ray emission of an early-type galaxy is an important tool for constraining the gas flow phase, especially in X-ray faint objects. Unfortunately, the amount of this contribution, which should be substantial, is still a matter of debate, due to the impossibility of estimating it directly from the available *Einstein* X-ray spectra (Forman, Jones, and Tucker 1985; Canizares et al. 1987). We know that a population of low mass X-ray binaries can explain the X-ray emission of the bulge of M31 (Fabbiano, Trinchieri, and Van Speybroeck 1987).

Progress in this field has come from a systematic investigation of the X-ray spectra of normal galaxies obtained by the *Einstein* satellite (Kim, Fabbiano, and Trinchieri 1992), and from the analysis of the ROSAT pointed observations of two early-type galaxies with low $L_X/L_B$ (Fabbiano, Kim, and Trinchieri 1994; see Fig. 1). On the average the X-ray emission temperature of early-type galaxies is found to increase with decreasing $L_X/L_B$, until the dominant contribution to the total emission comes from a in spirals, which can be fitted with bremsstrahlung emission at a temperature $kT > 3$ keV, and likely comes from binary X-ray sources (Kim et al. 1992). These authors also find a very soft thermal component at $kT \approx 0.2$ keV, in addition to the hard one, in a group of 6 early-type galaxies with the lowest values for $L_X/L_B$ [$29.3 < \log(L_X/L_B) < 30.0$, with $L_X$ in erg s$^{-1}$, $L_B$ in $L_\odot$, as throughout this paper]. ROSAT PSPC observations of NGC4365 and NGC4382, two galaxies belonging to the group above, confirm the results obtained with the *Einstein* data, and allow to define more accurately the spectral properties and luminosity of this very soft component (hereafter VSC). Its temperature is between 0.15 and 0.31 keV at the 90% confidence level, and its luminosity is half of the total $L_X$, in the best fit case, but large variations are allowed by the data (Fabbiano et al. 1994). Pellegrini and Fabbiano (1994) have investigated the origin of the VSC, in a hot interstellar medium and in stellar sources.

To investigate whether the VSC is a common feature of low $L_X/L_B$ early-type galaxies, and more generally to better understand the nature of their X-ray emission, we have analyzed the ROSAT PSPC pointed observation of NGC5866. *Einstein* observations showed that $\log(L_X/L_B) = 29.54$ for this galaxy, but it was not possible to perform a spectral analysis with the data (Fabbiano et al. 1992).

This paper is organized as follows: we present in Section 2 the main properties of NGC5866, in Section 3 the data analysis, in Section 4 we discuss the results and the implications for models of the X-ray emission of early-type galaxies, and in Section 5 we summarize the conclusions.

## 2. General characteristics of NGC5866

The main characteristics of NGC5866 are summarized in Table 1. This edge-on S0 galaxy belongs to the nearby group LGG396 (Garcia 1993), consisting of other two spiral galaxies besides NGC5866: NGC5907, the brightest of the group, and NGC5879. The velocity dispersion of the group is low (74 km s$^{-1}$, Huchra and Geller 1982), and no extended X-ray emission is detected, other than that associated with the galaxies (Forman et al. 1985).

Along the optical major axis of NGC5866 there is a prominent dust lane, extending for $\sim 60''$ (Takase, Kodaira, and Okamura 1984). This gas is possibly of external origin, in view of its misalignment by $5°$ with the optical body of the galaxy (Bertola, Buson, and Zeilinger 1992).

The gas (atomic and molecular) and dust content of NGC5866 is low, as is generally for S0 galaxies. Haynes et al. (1990) place an upper limit to the atomic HI mass of $3.2 \times 10^7$ $M_\odot$. Thronson et al. (1989) find extended CO emission, from which they deduce $3.2 \times 10^8$ $M_\odot$ of molecular gas. NGC5866 also belongs to the *IRAS* Bright Galaxy Sample (Soifer et al. 1987). The 12 $\mu$m emission seen by IRAS is likely due to photospheric and circumstellar emission from cool evolved red giant stars (Knapp, Gunn, and Wynn-Williams 1992). The far infrared emission $L_{fir} = 2.9 \times 10^{43}$ erg s$^{-1}$, calculated from IRAS data at 60 and 100 $\mu$m, is much lower than $L_B$, and typical of a normal galaxy; it is indicative of a very low star formation rate (David, Jones, and Forman 1992).

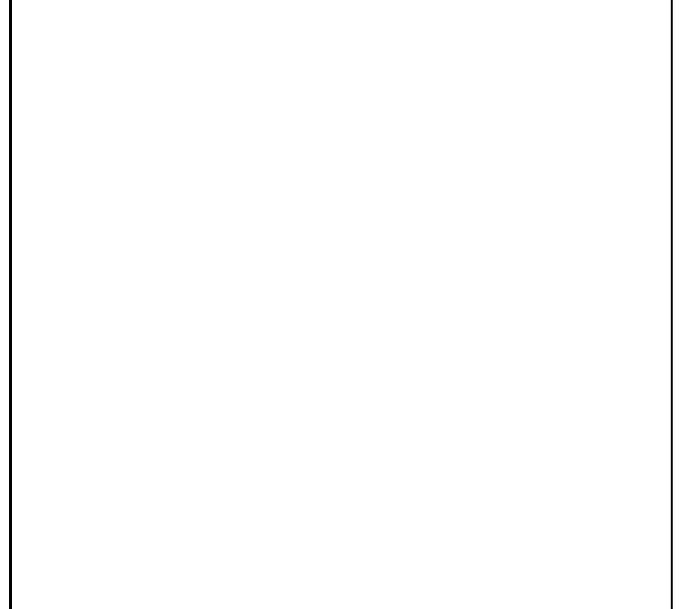

**Fig. 1.** The $L_X - L_B$ diagram for early-type galaxies with X-ray detection (data from Fabbiano et al. 1992). With filled symbols is plotted the group of 6 galaxies having a low $L_X/L_B$ value, considered by Kim et al. (1992), see Sect. 1. The NGC names are shown for the galaxy studied here, and those studied in Fabbiano et al. (1994). Also shown are two dashed lines, indicating the expected range for the stellar contribution to the X-ray emission (from Canizares et al. 1987).

Finally NGC5866 is a radio-quiet galaxy, since its radio luminosity is $5.9 \times 10^{27}$ erg s$^{-1}$Hz$^{-1}$ at 5 GHz (Fabbiano et al. 1987), and is confined within the optical body of the galaxy.

## 3. X-ray Data Analysis

NGC5866 was observed by the ROSAT PSPC (Trümper 1983, Pfeffermann et al. 1987) on 2–15 Dec 1991, for approximately 12300 seconds. We analyzed the data using the 'xray' package of IRAF, developed at SAO specifically for the analysis of *Einstein* and ROSAT X-ray data.

### 3.1. Spatial Analysis

In Fig. 2 a contour plot of the X-ray image is shown. The counts of the raw image have been taken, using data from pulse invariant gain-corrected (hereafter PI) spectral channels 11–240 ($\sim$0.1–2.4 keV). The PSPC data have been binned into 5″ pixels, and then smoothed with a gaussian of $\sigma = 15''$. An extended source is clearly visible, of extension larger than the effective ellipse, but smaller than the envelope ellipse – the first ellipse encircles half of the total optical light, the second one is the isophote of the 25 visual mag arcsec$^{-2}$ level (from Kodaira, Okamura, and Ichikawa 1990). The X-ray emission shows the same position angle of the optical emission; its extension is much larger than that of the previous *Einstein* image (Fabbiano et al. 1992).

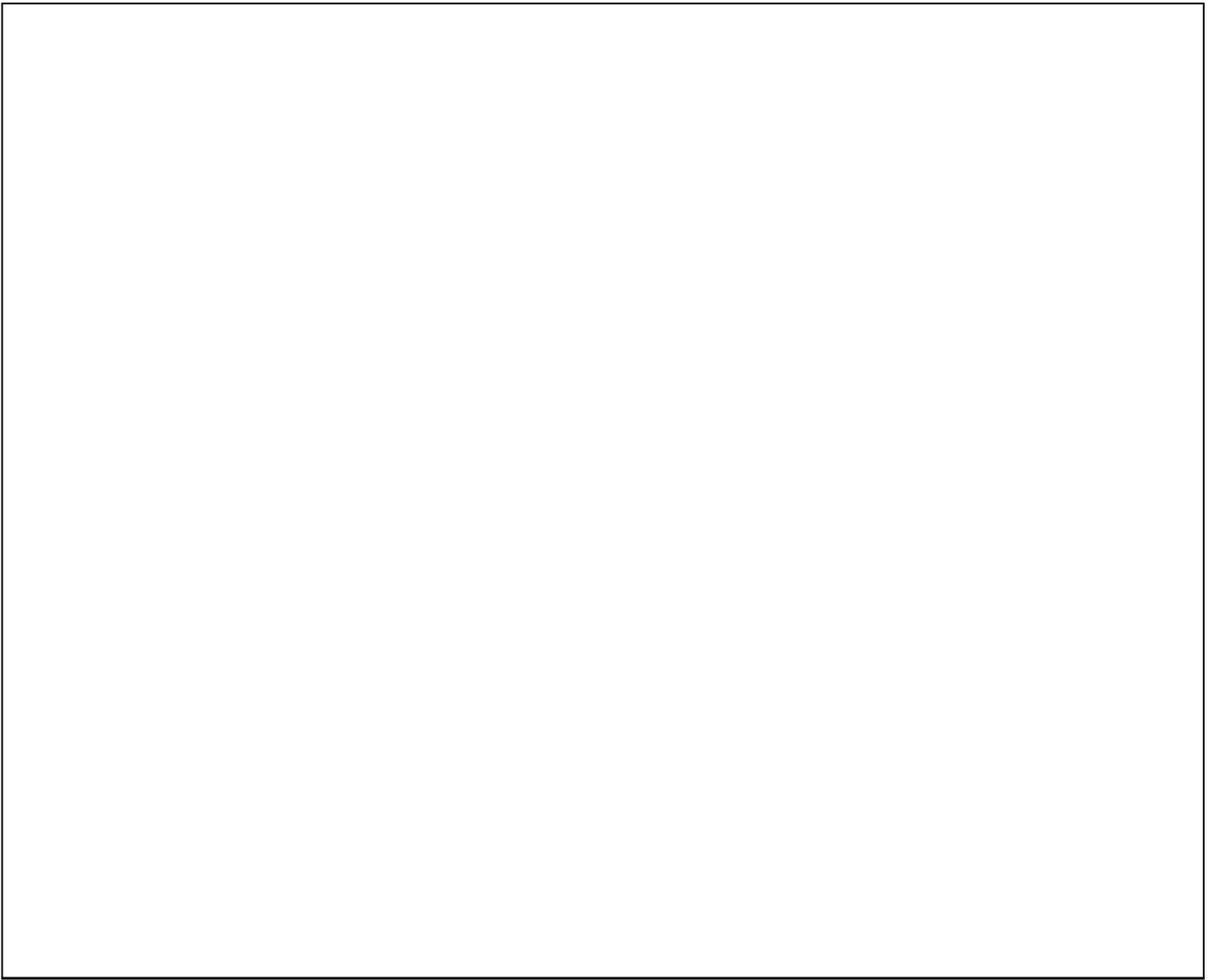

**Fig. 2.** Contour plot of the X-ray emission of NGC5866 (see Sect. 3.1). Contours plotted are 20, 25, 30, 35, 40, 50, 60, 75, 90 % of the peak intensity. The solid line ellipses indicate the effective and the envelope isophotes of the optical image (also the optical major axis is shown); the dashed ellipse indicate the extent of the X-ray image (see Sect. 3.1).

Table 1: General characteristics of NGC5866

| Type[a] | RA(J2000)[a] | Dec(J2000)[a] | d(Mpc)[b] | $L_B(L_\odot)$[b] | $N_H(cm^{-2})$[c] | $\sigma$(km/s)[d] | $v$(km/s)[d] |
|---|---|---|---|---|---|---|---|
| S0 | $15^h6^m29^s.36$ | $55°45'47''.71$ | 24.6 | $4.3\times10^{10}$ | $1.5\times10^{20}$ | 170 | 132 |

[a] from Kodaira, Okamura, and Ichikawa (1990)
[b] from Fabbiano, Kim, and Trinchieri (1992) (d=distance)
[c] Galactic neutral hydrogen column density from Stark et al. (1992)
[d] central stellar velocity dispersion and maximum rotational velocity of the stars, from Davies et al. (1983)

| Source Area[a] | Background Area[a] | Net Counts[b] | Exp. time (s)[c] | Rate (ct/s) |
|---|---|---|---|---|
| 237288 | 866171 | 387±32 | 12328±148 | 0.03 |

[a] Areas are in instrument pixels (1 pixel is 0.″5 wide); the region is an ellipse centered at 7725,7654 in the case of the source (see Sect. 3.1 for its size), and an annulus surrounding the source in the case of the background. This annulus has an internal radius of 300″, and is 100″ wide (a point source at 7080,7700 has been excluded from it).
[b] Background subtracted source counts, in the energy range (0.1–2) keV.
[c] Average exposure time in the source region, from the exposure map.

The X-ray isophotes show an elongation towards the south, where another emission region, which cannot be resolved as a separate source, is present. No objects are found to be falling in this region, after inspection of the file listing known sources in the image field, provided with the PSPC data.

To determine the size of the region from which to extract the counts for the spectral analysis, we need to determine precisely the X-ray source extent. We consider this to be the point at which the X-ray surface brightness of the source flattens onto the background level. Because of the small source extent, the background can be estimated from an annulus 100″ wide, surrounding the source. This can be proved as follows (see also Fabbiano et al. 1994). In the annulus the effective exposure time is expected to differ from that in the source region, because of various effects (local variations in the PSPC quantum efficiency, rib/mesh obscuration, telescope vignetting, aspect variation, and exposure time variation)[1]. An exposure map, made by folding the calibration map through the aspect history of the telescope, gives the effective exposure at each point in the detector, corrected with respect to the effects described above. This map, generated by the standard pipeline processing software, is included with the data set; it is based on a vignetting correction at 0.93 keV, which is appropriate for our purpose, since almost all of the background counts are below 1 keV, and the vignetting effect is constant up to 1 keV (Schlegel 1994). Using the exposure map, we derived the correction to be applied to the counts in the background region; on average this is 2.3%, and always less than 5%, so it is negligible with respect to the size of the uncertainties in the X-ray surface brightness profile.

[1] Particularly, vignetting effects become important at distances > 6′ off-axis (Snowden, Burrows, and Mendenhall 1994); this means they don't affect the source region, whose radial extension is clearly less than 4′ (see Fig. 2).

The background subtracted radial profiles along the major and minor axes of the source are plotted in Fig. 3a and 3b. These profiles are obtained considering two stripes, each one 30″ wide, one along the major and the other along the minor axis, and the PI channels 11–240. The profile along the major axis goes below the background level ∼180″ from the galaxy center. The situation is less clear along the minor axis, because the X-ray emission of NGC5866 merges with that from other sources in the south direction (Fig. 2). So, the source extent along the minor axis is derived from the X-ray surface brightness in the north direction; half minor axis turns out to be ∼ 110″ long, well comparable with the optical extension (Fig. 2).

Having determined the source extent, we can calculate the source net count rate, and extract the X-ray spectrum. The source net counts are estimated using the above determined X-ray ellipse (dashed line in Fig. 2), after removal of the portion of the diffuse emission at south that falls within it. Within the source region no other sources are visible by eye, and none are identified with the standard source detection routines. This results into 387±32 total net counts in the energy range (0.1–2.0) keV. By using the exposure map, we find that in the region covered by the source the effective exposure time is always within ±4% of an average exposure time of 12328 sec. So, in the energy range (0.1–2.0) keV, the net count rate is 0.03 counts/s (Table 2).

### 3.2. Spectral Analysis

The X-ray spectrum of the source has been derived extracting counts from the region described in Sect. 3.1, and subtracting the background as described there. The original spectral data, which come in 256 PI bins, are rebinned into 34 bins, covering an energy range from 0.07 to 2.48 keV, by the 'xray' package. The data are compared to models, after convolution with the instrumental and mirror response, via $\chi^2$ fits. In order to use the information contained in the channels which are inadequately filled for a $\chi^2$ fit statistic (Cash 1979), we have used both 'xray' and XSPEC, the X-ray spectral fitting package provided by the NASA/HEASARC, with which it's possible to group the original bins as desired. The channels corresponding to energies < 0.11 keV have been excluded from the fitting, because in this spectral region the response function of the PSPC is less well determined (Hasinger et al. 1992). The spectral response matrix released in January 1993 has been used in the fitting process; this is thought to be the best available for PSPC data taken after October 14, 1991.

| Model | $N_H(\mathrm{cm}^{-2})$ | $kT(\mathrm{keV})$ | 90% conf. range[a] | $\chi^2$ | $\nu$[b] | $L_X(\mathrm{erg/s})$[c] | 90% conf. range[a] |
|---|---|---|---|---|---|---|---|
| Bremsstrahlung | $1.3\times10^{20}$ | 0.7 | 0.5–1.2 | 11.8 | 15 | $2.8\times10^{40}$ | $(2.2$–$3.0)\times10^{40}$ |
| Raymond+Raymond | $1.1\times10^{20}$ | | | 11.5 | 13 | $2.3\times10^{40}$ | $(1.9$–$3.1)\times10^{40}$ |
| soft component | | 0.15 | < 0.22 | | | $1.2\times10^{40}$ | $(0.8$–$2.4)\times10^{40}$ |
| hard component | | 1.0 | > 0.67 | | | $1.1\times10^{40}$ | $(0.9$–$1.4)\times10^{40}$ |

[a] the range has been calculated for $10^{20}$ cm$^{-2}$ < $N_H$ < $2\times10^{20}$ cm$^{-2}$
[b] number of degrees of freedom of the fit
[c] the X-ray luminosity is in the (0.1–2) keV band, and unabsorbed; its uncertainty is dominated by the uncertainties in the spectral parameters (statistical errors are 8%). The X-ray luminosity in the (0.2-4) keV band derived from *Einstein* data is $1.48\pm0.46\times10^{40}$ erg s$^{-1}$, where the error is statistical only (Fabbiano et al. 1992).

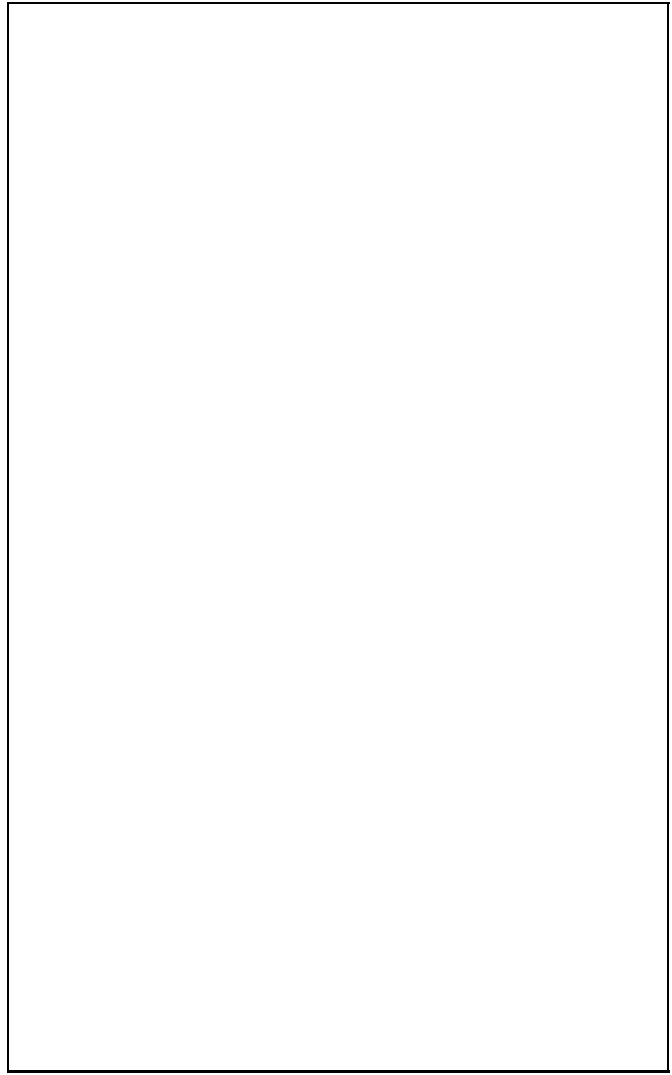

**Fig. 3.** Background subtracted X-ray surface brightness profile, along the major (a) and the minor (b) axis. The radius encircling 90% of the photons received by the combination of the ROSAT mirror and the PSPC is 0.45 arcmin at 1 keV (Hasinger et al. 1992).

The models used for the fits are the bremsstrahlung model, and the Raymond model (updated 1992 version of Raymond and Smith 1977) describing the thermal emission of an optically thin hot plasma, both from continuum and lines; the abundance ratios of the heavy elements are fixed to the solar values, because the limited spectral resolution and the large error bars in the spectral data don't allow to constrain them. These models have been used widely in previous works to describe the X-ray spectra of early-type galaxies (see Sect. 1). The absorption of the X-ray photons due to intervening cold gas along the line of sight is also applied, using the Morrison and McCammon (1983) cross sections.

The results of the spectral analysis are presented in Table 3. We first tried to fit single models to the data. The Raymond model with solar abundance is not a good representation of the data, because the probability of exceeding $\chi^2_{min}$ is just 2%, for a fit with 15 degrees of freedom. The bremsstrahlung model gives a better fit, the probability of exceeding $\chi^2_{min}$ is $\sim 70\%$; at the best fit the emission temperature is 0.7 keV, and $N_H = 1.3\times10^{20}$ cm$^{-2}$, comparable to the value in Table 1. At 90% confidence level, the temperature varies in the range 0.5–1.2 keV. Notice that a single bremsstrahlung component is equivalent to a Raymond component with zero abundance for the heavy elements; in fact, a thermal model with metal abundance < 5% solar (and $kT \sim 0.7$ keV) gives a fit as good as that unacceptable when increasing the abundance.

Then we tried the coupling of two thermal components with solar abundance (Fig. 4). At the best fit their temperatures are 0.15 and 1.0 keV, and their fluxes comparable (Table 3). The probability of finding a larger $\chi^2$ is $\sim 60\%$, so there is not an improvement with respect to the single bremsstrahlung model, but this two-component interpretation is physically more meaningful (see Sect. 4). If we constrain $N_H$ to be comparable to the line-of-sight neutral hydrogen column density due to the Galaxy ($10^{20}$ cm$^{-2} < N_H < 2 \times 10^{20}$ cm$^{-2}$), the temperatures of the two components vary in the ranges $0.08 - 0.22$ keV, and $> 0.67$ keV at the 90% confidence level (Fig. 5). At the same confidence level the soft/hard ratio can vary from 0.6 to 2.5. If one of the thermal components is described by a bremsstrahlung model, at the best fit we have a situation close to the fit with a single bremsstrahlung component: the Raymond component has a very low flux ($\sim 10^{39}$ erg s$^{-1}$), and the temperature of the bremsstrahlung component is 0.7 keV. The same happens when fitting with two bremsstrahlung components. Finally, trying to fit the data with heavy element abundances other than solar – from 10% to 3 times solar – does not give significant differences in the $\chi^2$ and the fit temperatures, due to the large uncertainties in the spectral data.

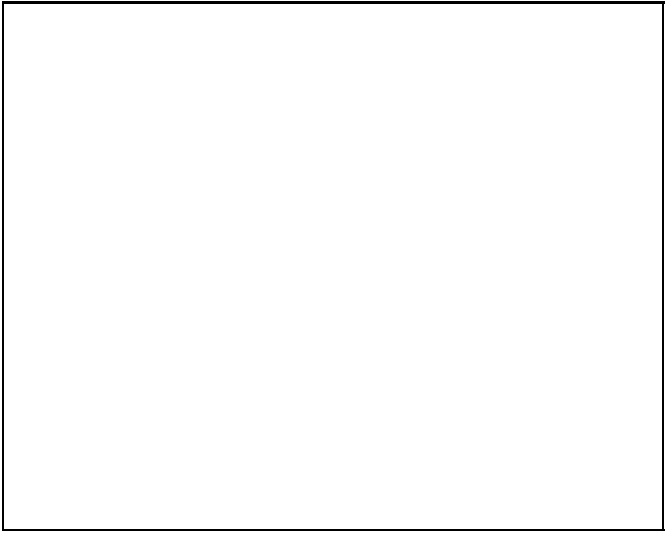

**Fig. 4.** The observed distribution of the PSPC spectral counts (symbols with error bars), in the 34 bins representation. The superposition of two Raymond components at the best fit is shown with a dashed line.

In Table 3 the unabsorbed luminosities in the 0.1–2 keV range are also given. These are slightly larger than those obtained from *Einstein* data (see note c in Table 3); the difference can be explained with a difference in the energy bands of sensitivity for the two satellites, and in the were detected by *Einstein*, and a 1 keV Raymond thermal emission had been assumed (Fabbiano et al. 1992). In fact, fitting the PSPC spectrum with such a model, gives just $L_X = 1.26 \times 10^{40}$ erg s$^{-1}$, a value in agreement with that of Fabbiano et al. , within the statistical uncertainties on the *Einstein* count rates.

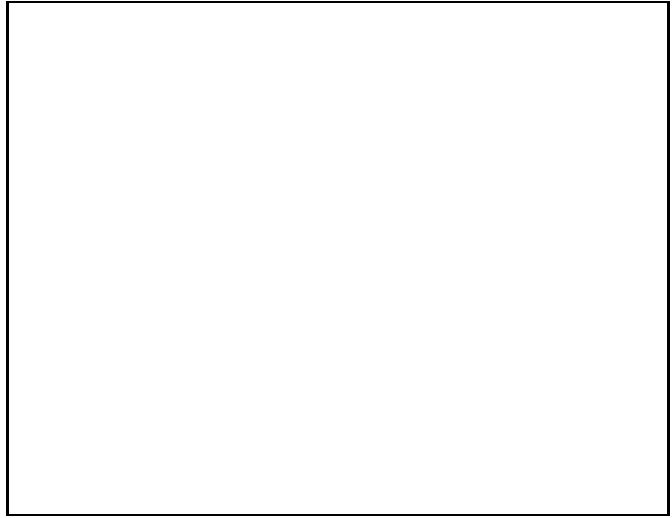

**Fig. 5.** The confidence contours corresponding to 68, 90 and 99% confidence level, for the temperatures of the two-component fit (the increments above the $\chi^2_{min}$ correspond to two interesting parameters, Avni 1976).

## 4. Discussion

Summarizing the findings of Sect. 3, acceptable models for the X-ray emission of NGC5866 are either a thermal component at a temperature between 0.5 and 1.2 keV with metal abundance $< 5\%$ solar, or two thermal components with solar abundance, the softer at a temperature around 0.1–0.2 keV, and the harder $> 0.6$ keV. The temperature of the hard component cannot be constrained for high values – the contours are open towards high values in Fig. 5 – because of the drop of the combined mirror and PSPC spectral response at high energies.

### 4.1. Origin of the X-ray emission of NGC5866

Our results are similar to the recent findings of Fabbiano et al. (1994), coming from the analysis of the PSPC data of the low $L_X/L_B$ galaxies NGC4365 and NGC4382 (see Sect. 1). Fabbiano et al. fit the spectral data either with a single temperature thermal model of $kT = 0.6 - 1.1$ keV, with abundance $<4\%$ solar, or with two thermal components of solar abundance, one at $kT = 0.15 - 0.31$ keV, the other at $kT > 1$ keV, contributing by comparable amounts to the total flux. They adopt the two-the data, since the one-component model would require either the not plausible discovery of a new class of X-ray emitting stellar sources, or the presence of a hot interstellar medium (hereafter ISM) at an extremely low abundance. Such a low abundance is not expected in an ISM originated by stellar mass loss (Renzini et al. 1993, Serlemitsos et al. 1993), unless this medium is primordial material recently accreted from outside the galaxy, which has also been swept of most of its own ISM. Adopting a two-temperature model, Fabbiano et al. suggest for the origin of the hard spectral component the integrated emission of low mass X-ray binaries, similar to that responsible for the bulge emission of early-type spirals (Fabbiano 1989). The possible origins of the VSC are discussed by Pellegrini and Fabbiano (1994), who have investigated whether this component can be due to a hot ISM, or is most likely originated by the collective emission of very soft stellar sources. As stellar soft X-ray emitters they have considered late-type stellar coronae, supersoft sources discovered by ROSAT in the Magellanic Clouds and M31 (e.g., Kahabka 1994), and RS CVn systems. All these candidates together could substantially explain the very soft emission, though none of them, taken separately, plausibly accounts entirely for its properties. The alternative origin of the VSC in the ISM has been investigated through hydrodynamical simulations of the secular behavior of gas flows, specific to the low $L_X/L_B$ galaxies NGC4697 and NGC4365. In the simulations the amount of dark matter and the rate of explosion of type Ia supernovae (hereafter SNIa) have been varied; these two ingredients, together with the depth of the stellar potential well which is measured by $\sigma$, determine the emission temperature of the hot gas. The results show that the gas temperature can be comparable to that of the VSC only in galaxies with quite shallow potential wells, i.e., with $\sigma \lesssim 200$ km s$^{-1}$, while for example $\sigma = 262$ km s$^{-1}$ in NGC4365. So, a hot ISM cannot be a general explanation of the very soft emission, although it can give a contribution to it.

The most likely interpretation of the X-ray data of NGC5866, and the related question of the presence of a hot ISM in a detectable amount, are discussed in the following. 1) In the one-component model, the source of the emission would be a hot ISM accreted from outside. This hypothesis is unlikely for various reasons: the presence of an intragroup medium has not been revealed (see Sect. 2); the temperature of this thermal emission (see Table 3) is quite high in comparison to that expected in NGC5866, due to its low $\sigma$ value, as estimated from models (Pellegrini and Fabbiano 1994); finally, in this case the hard emission from low mass X-ray binaries shouldn't show at all in the spectrum, at variance with what found already for galaxies of low $L_X/L_B$ ratio (see Sect. 1). 2) In the two-component model, the hard component cannot be produced by a hot ISM, for the same two last reasons of 1), and so its most likely explanation is the integrated emission of low mass X-ray binaries; the origin of the VSC is more a matter of debate: since $\sigma$ is just 170 km s$^{-1}$ in NGC5866, the average emission temperature of its hot ISM can be as low as $\sim 0.2$ keV, in the extreme case of a SNIa rate equal to the lowest values observed, and a low dark matter content[2] (Pellegrini and Fabbiano 1994). This point is further discussed below.

### 4.2. The X-ray emission of S0 galaxies

Although a detailed modeling of the structure of NGC5866, and hydrodynamical simulations of the hot gas behavior specific for this galaxy, are required for a definite answer, we can try to make some predictions concerning which gas flow phase is most likely on the basis of energetic arguments. Following Ciotti et al. (1991), we can estimate whether the global energy balance of the hot gas allows it to escape the galaxy by evaluating the ratio $\chi$ between $L^-_{grav}$ (the power required to steadily extract from the galaxy the stellar mass loss) and $L_\sigma + L_{SN}$, where $L_\sigma$ is the power made available to the gas by the thermalization of the stellar random motions, and $L_{SN}$ is the supernova heating per unit time. Assuming a standard mass model for NGC5866, i.e., a King profile for the luminous mass distribution and a quasi-isothermal dark halo (e.g., Ciotti et al. 1991), the previous ratio $\chi$ goes from $0.11/\vartheta_{SN}$, to $0.14/\vartheta_{SN}$, to $0.37/\vartheta_{SN}$, as the ratio of dark to luminous matter goes from 0 to 1 to 9 ($\vartheta_{SN}$ is the fraction of the rate of explosion of SNIa with respect to that estimated by Tammann (1982), and more recent observations place it between 0.25 and 1.1, e.g., Renzini et al. 1993). So, unless the dark matter is largely dominating over the luminous one *and* the SNIa rate is close to the lowest estimated values, the hot ISM is expected to be in a global wind.[3]

Since NGC5866 is considerably rotating, a wind regime could be even more likely than predicted above. In a more general frame, it is worth investigating the role of rotation to find an explanation of the recent finding that S0 galaxies tend to have lower X-ray luminosity per unit optical luminosity than do ellipticals (Eskridge, Fabbiano, and Kim 1994). The following arguments suggest that the role of rotation could be important. We can generalize the expression of $\chi$ introducing in it $L_{rot}$, the rotational power of the hot gas due to the streaming velocity of the stars around the galactic center ($v$ in Table 1): so now a reduction of the $\chi$ value up to 25%, for a fixed mass distribution, going from pressure supported to rotationally supported systems. Observationally at most $v/\sigma \approx 1.5$ (e.g., Busarello et al. 1992), with which the reduction in $\chi$ is as large as the 17%.[4] This is still significant, because hydrodynamical simulations show that small variations (within 10%) of $\chi$ when it is close to 1 cause large flow regime variations, and observational properties suggest that indeed $\chi \sim 1$ for most of the galaxies (Ciotti et al. 1991).

---

[2] Under these conditions the hot gas typically develops a *partial* wind; in this case an amount of gas sufficient to produce an $L_X$ comparable to that of the VSC can be retained by the galaxy.

[3] The predictions above involve a global energy balance, and so are not strictly valid in the case of a partial wind, which is actually the only flow phase in which the hot gas could have the emission temperature and the luminosity required to give origin to the VSC (see note 1). Previous hydrodynamical simulations however suggest that this requires a SNIa rate $\lesssim$ the lowest estimated values (Pellegrini and Fabbiano 1994).

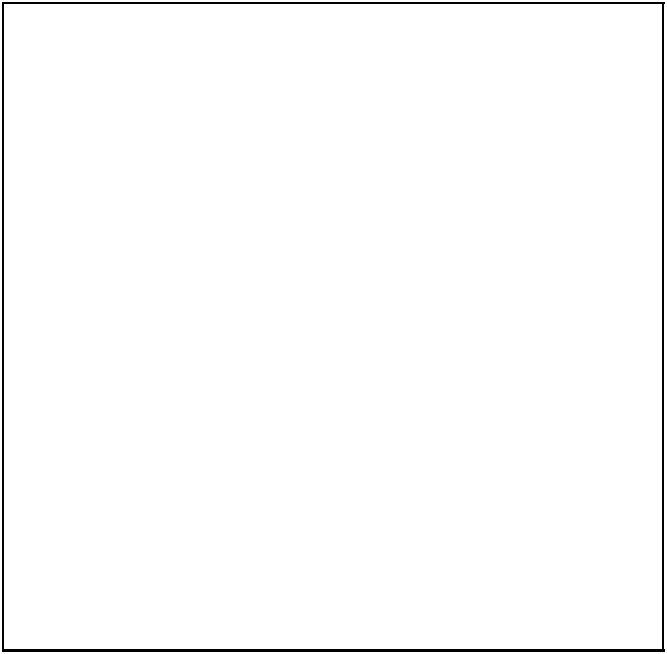

**Fig. 6.** The $L_X/L_B - v/\sigma$ diagram for S0 galaxies with X-ray detection. X-ray and blue luminosities are from Fabbiano et al. (1992), $v$ and $\sigma$ measurements are from Dressler and Sandage (1983), Davies et al. (1983), Whitmore et al. (1985), Bender et al. (1992), Busarello et al. (1992).

Using the available measurements of $\sigma$ and $v$, it turns out that the $L_X/L_B$ ratio, which is a measure of the hot gas presence, is anti-correlated with $v/\sigma$, and this supports our suggestion that the rotation could be at the origin of the systematic underluminosity in X-rays of S0s (Fig. 6). We are currently planning observations to obtain measurements of $v$ and $\sigma$ for the whole sample of S0s with X-ray detection; we are also investigating the problem with models for the internal dynamics of S0s, and two-dimensional hydrodynamical simulations of the hot gas flows behavior. Alternative explanations of the X-ray underluminosity of or a smaller $L_{grav}^-$, at fixed $L_B$, with respect to ellipticals.

---

[4] These reductions are calculated assuming no dark matter, otherwise they are larger; for example, a reduction up to 50% is allowed by the virial theorem (again for a fixed mass distribution), if the dark matter dominates over the luminous one.

### 4.3. The nature of the X-ray emission in low $L_X/L_B$ early-type galaxies

The first three ROSAT pointed observations of low $L_X/L_B$ objects allow to outline a more exhaustive picture of the nature of the X-ray emission across the $L_X$-$L_B$ plane of early-type galaxies. In fact, notwithstanding all the uncertainties discussed above (Sect. 4.1 and Sect. 4.2), the results obtained so far at least indicate that *even at high optical luminosities* a hot ISM doesn't dominate the X-ray emission of low $L_X/L_B$ galaxies (see for example the case of NGC4382 in Fig. 1). So, if a transition $L_B$, above which the $L_X$ is always dominated by a hot ISM, exists, its value must be higher than $\sim 10^{11}$ $L_\odot$. A transition luminosity quite lower than this had been suggested by Forman et al. (1985), and David, Forman and Jones (1991). David et al. for example predict, from their evolutionary models of the hot ISM, that with a SNIa rate equal to 0.25 that of Tammann (1982), and a massive dark halo, the critical blue luminosity is $L_B \sim 10^{10}$ $L_\odot$; moreover, according to these authors, galaxies brighter than $\sim 2 \times 10^{10}$ $L_\odot$ should always host a cooling flow (e.g., Sarazin and White 1988). The scatter in the X-ray luminosity at fixed optical luminosity would then be explained by a combination of large variations in the intrinsic properties of the galaxies (e.g., the stellar mass loss rate) and environmental effects. This scenario also implies that the stellar contribution to the X-ray emission is quite lower than that shown in Fig. 1.

An important contribution from stellar sources in low $L_X/L_B$ objects even at high optical luminosities is instead predicted naturally in the following scenario (see Ciotti et al. 1991): at every $L_B$ the scatter in $\sigma$ ($\sim$35% in the Faber-Jackson relation), or galaxy to galaxy differences in the amount of rotation, dark matter content, mass distribution, etc., give rise to a dispersion of the $\chi$ values around 1 (see Sect. 4.2). As a consequence a variety of flow phases can result: wind, partial wind, subsonic outflow and inflow. Galaxies hosting a wind are continuously swept of their hot ISM, which allows the stellar emission to become the dominant source of X-rays, and to be in the range estimated in Fig.1. Galaxies hosting either outflows or partial winds can accumulate some hot ISM, of $L_X \gtrsim$ that of stellar sources, while in those galaxies hosting an inflow the X-ray emission of the hot ISM definitely dominates thus masking the stellar sources contribution.

## 5. Conclusions

The nature of the X-ray emission of low $L_X/L_B$ early-type galaxies is becoming clearer after the first ROSAT PSPC pointed observations of these objects: NGC4365, NGC4382, and NGC5866. Here the case of the S0 galaxy

been obtained:

- The X-ray properties of NGC5866 are similar to those of the other two galaxies, including the presence of very soft emission, which seems to be a distinct characteristic of low $L_X/L_B$ objects. So, these can be recognized as a group with homogeneous properties, and a more exhaustive picture of the nature of the X-ray emission across the $L_X$–$L_B$ plane can be outlined.

- The most likely description of the X-ray emission mechanism consists of a very soft and a hard thermal component, of comparable flux. The origin of the hard component can be placed in low mass X-ray binaries, which also dominate the X-ray spectrum of bulges of spirals. The origin of the VSC could be placed in a hot ISM, due to the shallow potential well of NGC5866, or in stellar sources.

- Arguments are presented supporting a stellar origin for the VSC. These include the requirement of a SNIa rate equal to the lowest estimated values, for some hot gas with the needed emission temperature to be retained by the galaxy, and some energetic arguments considering also the role of rotation.

- It appears more likely that the hot ISM of NGC5866 is being swept by an internal mechanism such as a global wind, powered by SNIa's. An external mechanism such as ram pressure stripping is not possible instead, as this galaxy is a member of a small group with low velocity dispersion for which no intragroup medium has been detected.

- Whatever the origin of the VSC (entirely stellar, or, at least partially, in a hot ISM), ROSAT observations performed so far indicate that even at high optical luminosities a hot ISM doesn't dominate the X-ray emission of low $L_X/L_B$ galaxies. This supports the idea that a varying amount of hot gas, i.e., a different dynamical phase for the gas flows, ranging from inflow to subsonic outflow, to partial wind, to global wind, accounts for the scatter in the $L_X$–$L_B$ diagram of early-type galaxies.

- The fact that S0 galaxies are on average less X-ray luminous than elliptical galaxies, at the same optical luminosity $L_B$, could be explained with a higher rotation of S0 with respect to ellipticals, which makes easier for the hot gas to escape the galaxy. Alternatively a higher rate of SNIa's could be invoked, or a shallower potential well, compared to ellipticals of the same $L_B$.

Among the most important questions left open are that of the precise normalization of the stellar contribution to the X-ray emission of early-type galaxies, and that of the origin of the very soft emission. The first point will be investigated further thanks to ROSAT pointed observations of other X-ray faint galaxies, so that a statistically significant sample will be collected. ASCA observations, with a higher energy resolution and a larger spectral band, will help evaluate especially the contribution of hard binaries. This will be useful also for a more accurate determination of the hot gas temperature in X-ray bright galaxies, hard stellar emission has been made. The second point can be addressed investigating whether the fraction of very soft emission varies significantly from galaxy to galaxy, or whether its temperature depends on $\sigma$; this, which could point towards a hot ISM origin, seems not to be the case after the analysis of the first three pointed observations of low $L_X/L_B$ galaxies.

*Acknowledgements.* I thank L. Ciotti, A. D'Ercole, G. Fabbiano, P. Goudfrooij, and A. Renzini for discussions and suggestions.